\documentclass[prl,aps,twocolumn,groupedaddress,floats,final,superscriptaddress,nopacks]{revtex4-2}
\usepackage{graphicx}
\usepackage[utf8x]{inputenc}
\usepackage{lineno}
\usepackage{color}
\usepackage{subfigure}
\usepackage{latexsym}
\usepackage{epsfig}
\usepackage{psfrag}
\usepackage{amsmath,amsfonts,amssymb,bm,ulem}
\usepackage{float}
\usepackage{ulem}
\usepackage{braket}
\usepackage{hyperref}
\usepackage{tikz}
\usepackage{pifont}
\usepackage{xcolor}
\usepackage{microtype}

\newcommand{\vk}{{\mathbf{k}}}

\begin{document}

\title{Variational Diagrammatic Monte-Carlo Built on Dynamical Mean-Field Theory}
\author{Yueyi Wang}
\affiliation{Center for Materials Theory, Department of Physics and Astronomy, Rutgers University, Piscataway, New Jersey 08854, USA}
\author{Kristjan Haule\textsuperscript{\textdagger}}
\affiliation{Center for Materials Theory, Department of Physics and Astronomy, Rutgers University, Piscataway, New Jersey 08854, USA}
\begin{abstract}
  We develop a variational perturbation expansion around dynamical mean-field theory (DMFT) that systematically incorporates nonlocal correlations beyond the local correlations treated by DMFT. 
  We apply this approach to investigate how the DMFT critical temperature is suppressed from its mean-field value and how the critical behavior near the finite-temperature phase transition evolves from the mean-field to the Heisenberg universality class. 
  By identifying the symmetry breaking of paramagnetic diagrammatic expansions as a signature of the Néel transition, we accurately predict the Néel temperature of the three-dimensional cubic Hubbard model across all interaction strengths with low computational cost. 
  Introducing a variational order parameter, our method can be applied to both paramagnetic and long-range ordered states, such as antiferromagnetic order. 
  We compute magnetization and antiferromagnetic susceptibility, demonstrating minor corrections to DMFT solutions in the weak-coupling regime while revealing significant modifications to these properties in the intermediate correlation regime. 
  From the analysis of critical exponents, we establish the emergence of Heisenberg critical behavior beyond the mean-field nature of DMFT.
\end{abstract}

\maketitle

\textbf{Introduction.}
The computational design of materials has emerged as a cornerstone of modern condensed matter physics and materials science. Established methods, such as density functional theory (DFT), have enabled the prediction of millions of theoretically stable compounds, expanding the landscape of potential materials~\cite{Merchant2023}. These simulations also play a pivotal role in guiding experimental efforts in materials discovery~\cite{Szymanski2023}. 
A persistent challenge, however, lies in accurately predicting physical properties and excited states, a limitation that constrains the broader impact of these theoretical tools.
For instance, predicting the superconducting critical temperature ($T_c$) of unconventional superconductors remains beyond reach. While cluster Dynamical Mean Field Theory (DMFT) offers an ab-initio framework for simulating unconventional superconductivity~\cite{Benjamin2024,DCAreview,CDMFTSC,NONLOCALDMFTREVIEW,Lieb_DMFTSC}, this method can provide only the mean-field estimate of $T_c$, that can be very inaccurate. For example, single-site DMFT predicts a Curie temperature of $T_c = 1552\,$K for elemental iron~\cite{PhysRevLett.120.187203}, in stark contrast to the experimental value of $1043\,$K. Addressing these discrepancies requires methodological advances to refine ($T_c$) predictions, thereby enhancing the reliability of DMFT and related approaches in materials design. Such progress is essential for leveraging computational tools to their full potential in advancing materials science.

In this letter we develop a variational diagrammatic Monte-Carlo (VDMC) method
that can be combined with the DMFT to compute the transition temperature and critical phenomena in the ordered state very accurately. We develop and test the method on the three dimensional Hubbard model in the antiferromagnetic phase, but we stress that the method is generic and can be applied to other phase transitions, including to the unconventional superconductors. Most importantly, the method can predict $T_c$ across all interactions strengths, from the weak to the strong coupling, even though it has simplicity of the weak coupling expansion.

To benchmark our method we use the Hubbard model in three dimensions at half-filling
\begin{align}
  \mathcal{\hat{H}}=-t\sum_{ \langle i,j\rangle \sigma} \hat{c}^{\dagger}_{i \sigma} \hat{c}_{j\sigma}+U\sum_i \hat{n}_{i \uparrow}\hat{n}_{i \downarrow}
  -\mu \sum_{i \sigma}\hat{n}_{i\sigma},
  \label{Hubbard}
\end{align}
in which the antiferromagnetic transition temperature has been determined quite precisely by variety of methods~\cite{DDMCTN,DGA2,DCATN,TUFRGTN,DFandDMFTTN,QMCTN,secondorderlocalapprox,DGA3,partialsum,NONLOCALDMFTREVIEW,DGAattractive,THERMOLIMIT}.
Here $t$ is the hopping amplitude between nearest neighbours (taken as the unit of energy), $U$ the strength of local Coulomb interaction, and $\mu$ is the chemical potential. 
$\hat{n}_{i \sigma}=\hat{c}^{\dagger}_{i \sigma}\hat{c}_{i\sigma}$ is the particle number operator.

DMFT~\cite{DMFTreview} correctly predicts the metal–insulator transition in the 3D Hubbard model~\cite{DMFT} and provides the correct transition temperature scale. However, neglecting nonlocal correlations places the critical behavior in the mean-field rather than the Heisenberg universality class. Extensions such as the dual-fermion approach~\cite{DFandDMFTTN}, the dynamical vertex approximation (D$\Gamma$A)\cite{DGA}, and cluster DMFT\cite{DCATN,DMFTMATreview} partly remedy this but are computationally demanding and sometimes biased. For example, ladder-type D$\Gamma$A enforces the local spin sum rule via a Moriya-$\lambda$ correction, analogous to two-particle self-consistent theory (TPSC)~\cite{TPSC}, and yields the same spherical-model criticality as TPSC in the attractive Hubbard model~\cite{DGAattractive}. Hence, an unbiased, controlled treatment of criticality remains urgently needed, along with more cost-effective nonlocal DMFT extensions for materials design.

On the other hand, perturbative expansions are typically reliable only in the weak coupling regime. Even when the propagators are dressed, these expansions can fail as soon as divergences appear in the two-particle vertex, a phenomenon that coincides with the multivaluedness of the Luttinger-Ward functional~\cite{PhysRevLett.119.056402,PhysRevB.97.125141}.
In the strong interaction regime, expansions around such bold Green's functions may even converge to an incorrect solutions~\cite{misconverge}, making the non-self-consistent expansions preferable because such wrongful convergence can not occur~\cite{HomotopicAction}.
However, the recent resurgence of perturbative expansions, known as diagrammatic Monte-Carlo (DiagMC)~\cite{diagMC}, have provided new perspectives on overcoming these challenges. 
DiagMC has enabled controlled summation of diagrams to high orders with remarkable accuracy in certain cases, such as the polaron problem~\cite{diagMCpolaron}, and even for the Hubbard model~\cite{DiagMCfermions} in the regime of moderate interaction strength.
The analysis of the perturbative series and its divergence near the second order phase transition was used to detect precise position of the Néel transition in the 3D Hubbard model\cite{CDETPMTN}.
DiagMC has also been combined with dual-fermion frameworks~\cite{DiagMCDF1,DiagMCDF2}, but these approaches break down near the DMFT N\'eel temperature, restricting access to the critical regime and preventing a controlled determination of $T_N$.

There is a growing body of works demonstrating how perturbative expansion can be regularized by choosing a starting point that incorporates the collective behavior of the system~\cite{UEG1,UEG2,Pseudogap,Rossi_2020,Hubbard22,BCSvariationalperturbation}. This approach, also called the homotopic action approach~\cite{HomotopicAction}, emphasizes the critical importance of selecting an appropriate starting point for the perturbative expansion—a highly nontrivial task with a rich history. Several early studies incorporated variational techniques into perturbative expansions~\cite{varpert1,varpert2,Kleinertsbook}, and more recent efforts have extended these ideas to the uniform electron gas~\cite{UEG1,UEG2} and to the Hubbard model~\cite{BCSvariationalperturbation,VarPertHF}. 
By combining DiagMC with variational perturbation expansions from the mean-field solution, significant progress has been made in predicting $T_N$ and other observables in the Hubbard model under weak-to-intermediate interactions~\cite{BCSvariationalperturbation,VarPertHF}. 
However, these approaches are ineffective in the intermediate-to-strong coupling, where simple mean-field starting points fail to adequately capture the strongly correlated physics.

To apply the idea of the homotopic action~\cite{HomotopicAction} in the strongly correlated regime, it is essential to choose a starting point that captures the local moment regime, since this state is inaccessible from the Fermi liquid via finite-order perturbation theory. For this reason, the DMFT self-energy serves as a vastly superior starting point. It provides qualitatively correct description of the normal state across a broad range of interaction strengths, encompassing both the Fermi liquid and the local moment regimes. This stands in stark contrast to previous implementations of the homotopic action with self-energy corrections that are insufficient for describing strongly correlated regimes. 
In this letter, we implement a variational perturbative expansion around the DMFT solution in the ordered state to provide qualitatively correct, non-perturbative starting point even at strong coupling. Using 
the convergence behavior of variational perturbation,
we develop an efficient method to determine the Néel temperature for $U/t<15$ with high accuracy. Furthermore, estimated from the optimal convergence of the variational perturbation, the critical behavior of magnetization and susceptibility reveal the shift from the mean-field to the Heisenberg universality class. 

\smallskip

\noindent \textbf{Method.}
Our method is most concisely formulated in terms of the effective action $S(\xi)$:
\begin{widetext}
\begin{eqnarray}
\label{rearrangedaction}     
  S(\xi) =  \sum_{\braket{ij}\sigma}\int_0^\beta\int_0^\beta c_{i\sigma}^\dagger(\tau)\left[
  \delta(\tau-\tau') \left(\delta_{ij}( \partial_\tau -\mu+\frac{1}{2}\sigma^z_{\sigma\sigma}p_i)+t_{ij}\right)
  +\delta_{ij}\Sigma^{\mathrm{imp}}(\tau-\tau')\right]c_{j\sigma}(\tau')d\tau d\tau'
 \nonumber \\
  +\xi U \sum_i\int_0^\beta c^\dagger_{i\uparrow}(\tau)c^\dagger_{i\downarrow}(\tau)c_{i\downarrow}(\tau)c_{i\uparrow}(\tau) d\tau
  -\sum_{i}\int_0^\beta\int_0^\beta c_{i\sigma}^\dagger(\tau)  \left(\Sigma^{\mathrm{imp}}_{\xi}(\tau-\tau')
+\delta(\tau-\tau')\frac{\xi}{2}\sigma^z_{\sigma\sigma} p_i\right)
  c_{i\sigma}(\tau') d\tau d\tau' .
\end{eqnarray}    
\end{widetext}  
Here $U$ and $t_{ij}$ are parameters of the Hubbard model Eq.~\ref{Hubbard}. 
The second and third terms in Eq.~(2), which together constitute the interaction,  depend explicitely on the bookkeeping parameter $\xi$; the second terms represents the standard Coulomb interaction, while the third term serves as a conter-term.
In our perturbative expansion, we expand the action systematically in powers of $\xi$, treating both interaction terms on an equal footing. After generating the expansion, we set $\xi=1$ to ensure that the action coincides exactly with that of the Hubbard model.

The first term of the action defines a shifted non-interacting Green’s function
\begin{equation}
    G^0_{\sigma}(i\omega_n,\textbf{k})=[i\omega_n+\mu-\epsilon_{\textbf{k}}-\sigma^z_\sigma p_i/2-\Sigma^{\mathrm{imp}}(i\omega_n)]^{-1}
\end{equation}
where $\Sigma^{\mathrm{imp}}=  ( {\Sigma^{\mathrm{imp}}_{\uparrow}}+{\Sigma^{\mathrm{imp}}_{\downarrow}})/2$ is the spin-averaged DMFT self-energy. 
The exchange splitting is introduced variationally via $p_i=(-1)^i \alpha U$, which alternates on the two sublattices of the double unit cell ($i=0,1$ for the two sublattices);  $\alpha$ is the variational parameter. The reason for using variational exchange splitting is that DMFT overestimates exchange splitting in the ordered states, which slows the convergence of the series at strong coupling. By choosing a smaller exchange splittings in $G^0$ we can accelerate convergence substantially. We therefore determine 
$\alpha$ variationally and adopt the value 
$\alpha_{\mathrm{opt}}$ that yields the fastest convergence of the perturbative expansion.

The self-energy $\Sigma^{\mathrm{imp}}_{\sigma,\xi}$ entering the counterterm is written as a power series,
\begin{equation}
\Sigma^{\mathrm{imp}}_{\sigma\;\xi}=\sum^{\infty}_{n=1}\Sigma^{\mathrm{imp}(n)}_{\sigma}\xi^n
\end{equation}
where $\Sigma^{\mathrm{imp}(n)}_{\sigma}$ is the sum of all $n$-th order skeleton local diagrams 
constructed with the local DMFT Green’s function ($\sum_{\textrm{k}}1/(i\omega+\mu-\varepsilon_{\textbf{k}}-\Sigma^{\mathrm{imp}}_{
\sigma
}(i\omega_n))$) and interaction $U$. 
The spin-averaged counterpart, $\Sigma^{\mathrm{imp}}_{\xi}=\frac{1}{2}(\Sigma^{\mathrm{imp}}_{\uparrow\xi}+\Sigma^{\mathrm{imp}}_{\downarrow\xi})$ 
is used in the action whenever no spin label appears. Our algorithm systematically generates and evaluates every Feynman diagram up to order $\xi^n$ generated by Eq.~\ref{rearrangedaction}; the procedure is detailed in the first section of the Supplementary Information\cite{Supplementaryinfo}, which includes Refs.~\cite{DopedDMFT,dopedDGA,alpha_shift,ThermoQMC,AFQMCTN,Cvkink}.
A key feature of our approach is the distinction between the non-perturbative self-energy and its perturbative counterpart.
In the first line of Eq.~\ref{rearrangedaction}, which defines
$G^0_\sigma(i\omega,\textbf{k})$,
$\Sigma^{\mathrm{imp}}_{\xi=1}$  is obtained non-perturbatively with a quantum-impurity solver. In contrast, the counterterm employs the perturbative version $\Sigma^{\mathrm{imp}}_\xi$, which generates
the local skeleton diagrams only up to the order $\xi^n$. 
These counterterms precisely cancel the most divergent subset of (local) diagrams produced by the Coulomb interaction. This procedure guarantees that all—and only—Feynman diagrams beyond DMFT up to order $n$ are systematically generated and summed. Details about calculation of observables are discussed in the End Matter and Supplementary Information~\cite{Supplementaryinfo}.

\begin{figure}[t]
\centering
  \subfigure{\includegraphics[scale=0.5, trim=5mm 3mm 0mm 18mm, clip]{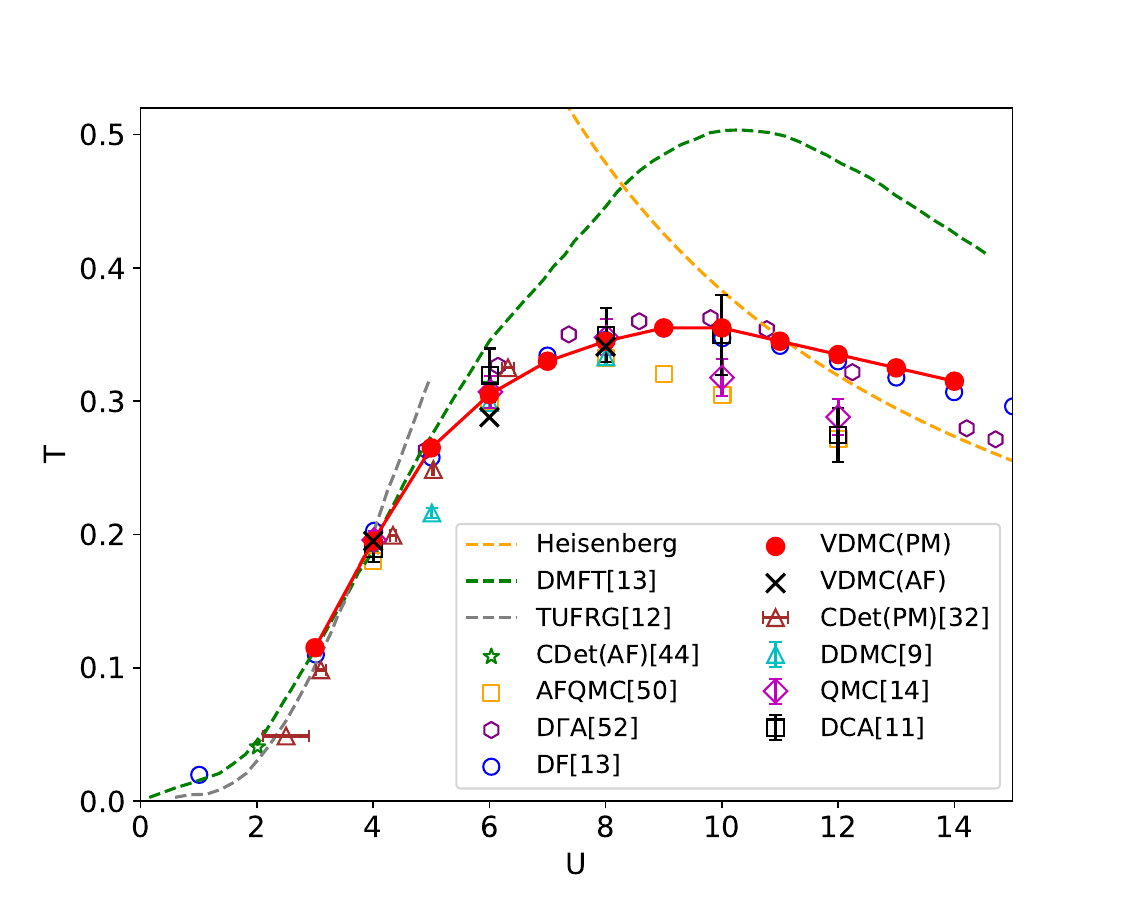}}
\caption{Comparison of the Néel temperature $T_N(U)$ obtained using VDMC with results from other numerical methods: the truncated-unity functional renormalization group (TUFRG)~\cite{TUFRGTN}; diagrammatic Monte Carlo methods, including CDet(PM)~\cite{CDETPMTN} and CDet(AF)~\cite{VarPertHF}; diagrammatic determinant Monte Carlo (DDMC)~\cite{DDMCTN}; auxiliary-field quantum Monte Carlo (AFQMC)~\cite{AFQMCTN}; quantum Monte Carlo (QMC)~\cite{QMCTN}; the dynamical vertex approximation (D$\Gamma$A)~\cite{DGATN}; the dynamical cluster approximation (DCA)~\cite{DCATN}; and the dual fermion method (DF)~\cite{DFandDMFTTN}. VDMC(PM) results are calculated using a small splitting starting point, whereas VDMC(AF) results are obtained by extrapolating the magnetization $m(T)$ to zero. Error bars for our results are smaller than the marker size.}
\label{Fig1}
\end{figure}

\noindent \textbf{Results and Discussions.} 
We first present our estimation of the Néel temperature ($T_N$) as a function of the interaction strengths $U$ in Fig.~\ref{Fig1}. Our results, labeled VDMC(PM) and VDMC(AF), are compared with those obtained by other numerical approaches~\cite{CDETPMTN,DCATN,DDMCTN,DFandDMFTTN,DGATN,AFQMCTN,QMCTN,TUFRGTN,VarPertHF}. 
In the weak interaction regime, the antiferromagnetism is primarily driven by the Slater mechanism~\cite{slaterAFM}, leading to an increase in $T_N$ as the interaction strength grows. It reaches the maximum value around $U\simeq 8\sim 10$, before decreasing toward the Heisenberg limit, where it is governed by the superexchange coupling $T_N\simeq 3.83 /U$~\cite{HeisenbergTN}.
Although it is widely believed that the strongly interacting regime is challenging to access with perturbative methods, 
our approach successfully determines the Néel temperatures across the entire interaction range. It provides significant corrections to DMFT and shows good  agreement with other advanced but more computationally intensive methods. 
This success can be attributed to (i) using DMFT self-energy instead of perturbative summation of self-energy diagrams to dress the non-interacted propagator as the starting point of perturbation  
(ii) allowing for long range order in the starting point, which regularizes the expansion.

We estimate $T_N$ in two different ways: The first is denoted by VDMC(PM) in Fig.~\ref{Fig1} (red dots) and is based on the perturbative properties at infinitesimal $\alpha$. We start by very small symmetry breaking in Eq.~\ref{rearrangedaction} ($\alpha\ll 1$), and then we check if higher order perturbative corrections restore the symmetry breaking, or, enhance it. If the chosen temperature is too high for the Néel order, 
small symmetry breaking in the starting point leads to magnetization which is reduced with increasing perturbative order. On the other hand, if the temperature is below the Néel point, a small splitting in the starting point leads to magnetization which is monotonically increasing with perturbative order, or in same cases diverges.
While offering a decent accuracy for estimation of the Néel temperature, this approach does not require one to estimate the critical behavior or physical magnetization of the system, works at arbitrary interaction strength, and is computationally cheap.  
More details on this method are provided in the End Matter and Supplementary Information~\cite{Supplementaryinfo}.
Predictions obtained using VDMC(PM), which extend into the strong-coupling regime, show the closest agreement with the dual fermion method~\cite{DFandDMFTTN} and remain reasonably close to D$\Gamma$A results~\cite{DGATN}. 
In the strong-coupling limit ($U > 15$), our method slightly overestimates $T_N$ because the series convergence slows and becomes nonmonotonic. Extending into this regime will require higher-order expansions, which we leave for future work.

\begin{figure}[t]
\centering
  \subfigure{\includegraphics[scale=0.55, trim=8mm 4mm 0mm 15mm, clip]{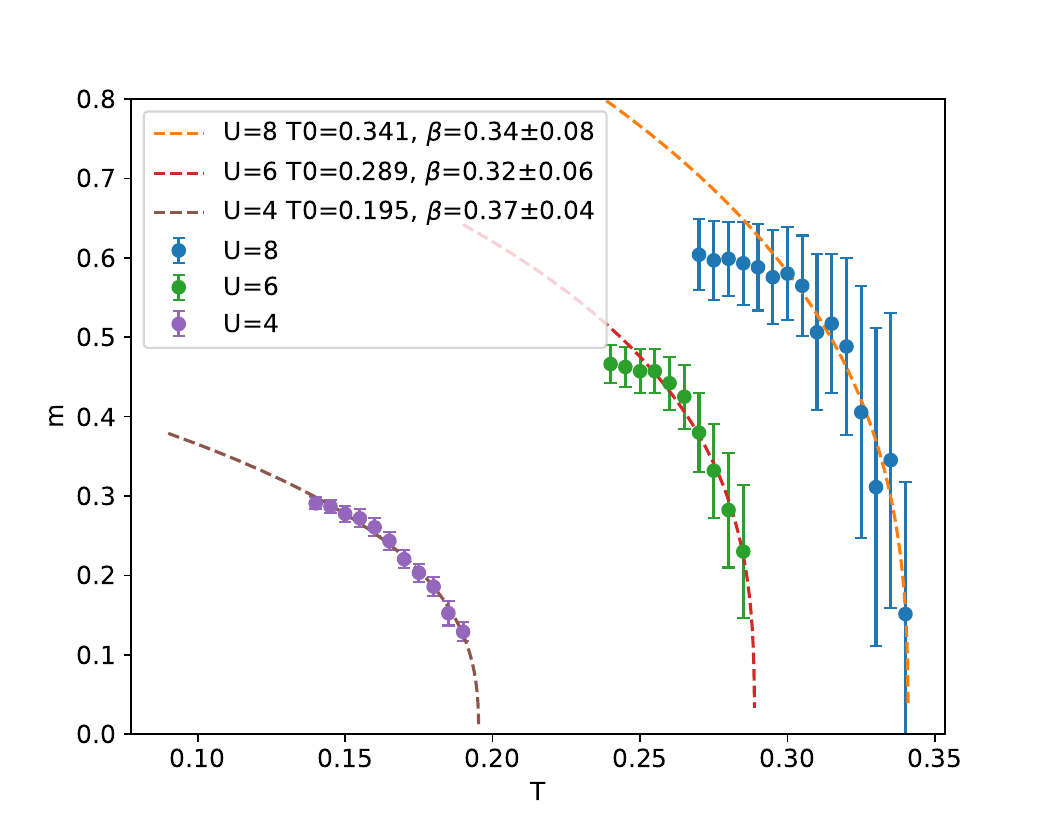}}
  \caption{
    Magnetization $m(T)$ and critical behavior at different interaction strengths.
    The dashed lines indicate the critical behavior fitting: $m(T)=A(T_N-T)^\beta$.}
\label{Fig2}
\end{figure}

The alternative estimation of the Néel temperature is obtained from extrapolating physical magnetization $m(T)$ to the temperature at which it vanishes (see Fig.~\ref{Fig2}). This lead to more precise estimation, but is computationally more expensive, and by using current algorithms, it does not work beyond the DMFT Mott transition.
We fit $m(T)$ with ansatz for critical behavior $m(T)=A(T_N-T)^\beta$ where the critical exponent is expected to be $\beta \approx 0.37$ for the $O(3)$ Heisenberg universality class~\cite{DGA2,DFandDMFTTN,HeisenbergExponents,QuantumCriticality}.
In practice, the fit yields $\beta(U=4)=0.37\pm0.04$, $\beta(U=6)=0.32\pm0.06$, $\beta(U=8)=0.34\pm0.08$, which are reasonably consistent with the expected value, and show significant correction from DMFT mean field exponent of $\beta=0.5$.
The extrapolation of magnetization to $m=0$ gives the second estimate of $T_N$, namely: $T_N(U=4)=0.20$, $T_N(U=6)=0.29$, $T_N(U=8)=0.34$, represented by VDMC(AF) in Fig.~\ref{Fig1}. 
These estimates are slightly lower than $\mathrm{VDMC(PM)}$ estimates and are in excellent agreement with other advanced numerical methods, indicating that the $\mathrm{VDMC(PM)}$ algorithm for $T_N$ is also accurate.

\begin{figure}[t]
\centering
  \subfigure{\includegraphics[scale=0.35, trim=0mm 0mm 0mm 0mm, clip]{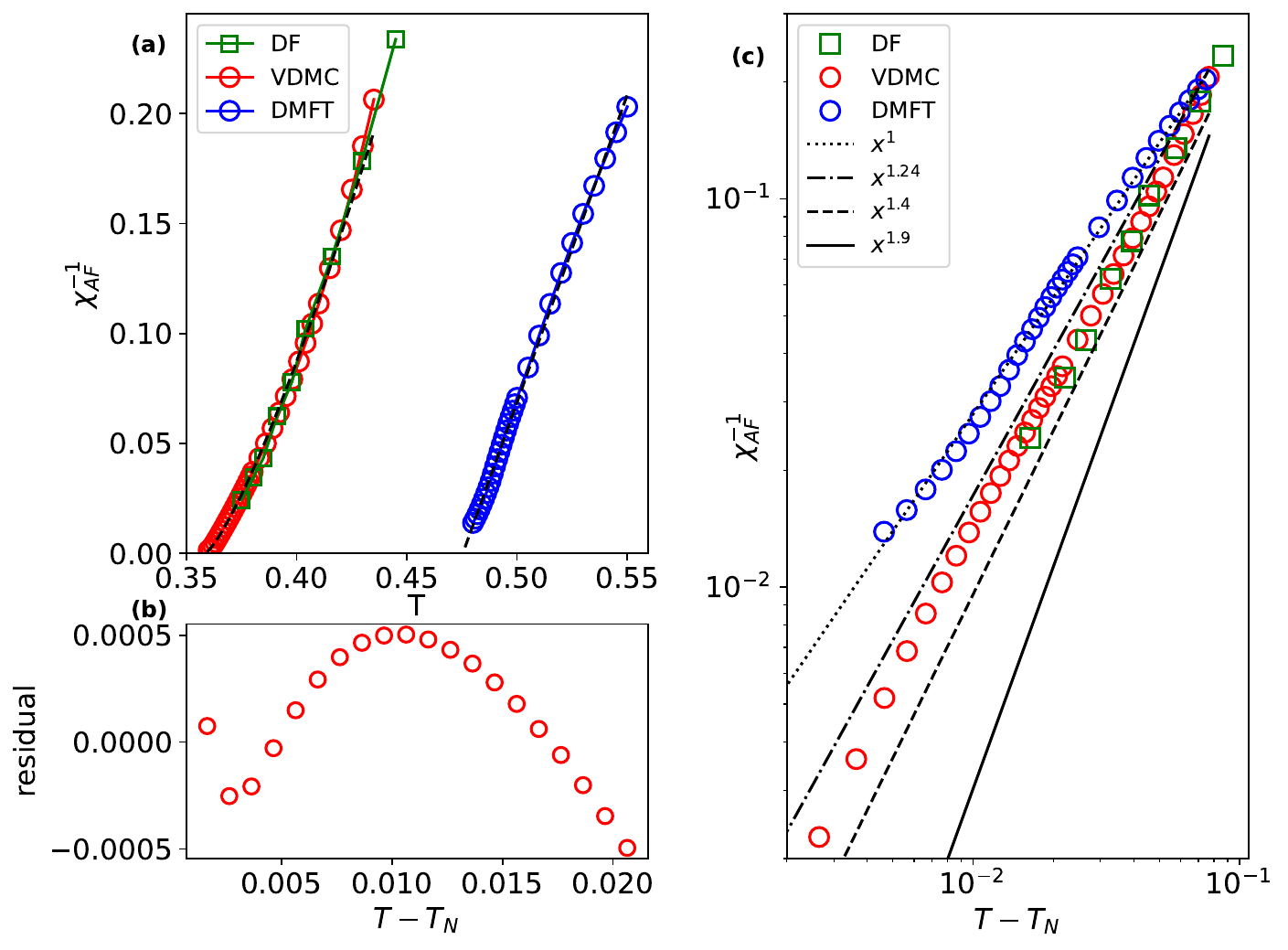}}
    \caption{Antiferromagnetic susceptibility $\chi_{AF}=\chi(\pi,\pi,\pi)$ at $U=11$ for DMFT, VDMC, and dual Fermion~\cite{DFandDMFTTN}. (a) DMFT, VDMC and dual Fermion $\chi_{AF}$ plotted on linear scale. The dashed lines show the fit to the form $\chi_{AF}=A(T-T_N)^{-\gamma}$. (b) The residuals of the VDMC $\chi_{AF}$ fit close to $T_N$, which shows details of fitting around the phase transition. (c) DMFT and VDMC $\chi_{AF}$ plotted on log-log scale, compared with dual Fermion results~\cite{DFandDMFTTN} and several power-laws: mean field $\gamma_{\mathrm{MF}}=1$, Ising $\gamma_{\mathrm{Ising}}=1.24$, Heisenberg $\gamma_{\mathrm{Heisenberg}}=1.4$, and $\gamma=1.9$ proposed in Ref.~\cite{DGAattractive}.}
\label{Fig3}
\end{figure}

To confirm that our approach gives the Heisenberg critical behavior, we calculate the antiferromagnetic susceptibility $\chi_{AF}=\chi(\pi,\pi,\pi)$. The data at $U=11$ is displayed in Fig.~\ref{Fig3}.
In DMFT, we compute the susceptibility from the response to a staggered external magnetic field, which scales as $\chi_{AF} \sim 1/(T-T_N)$. This behavior is consistent with earlier analytical results~\cite{Krien2018PhD} and calculations based on the two-particle vertex function~\cite{DMFTXAF}.

After applying perturbative corrections to DMFT, the critical temperature is reduced to $T_N = 0.358$, matching the value previously reported in the literature. The critical exponent is found to be $\gamma = 1.286$, which deviates significantly from the DMFT mean-field exponent ($\gamma = 1$) and the anomalously large exponent ($\gamma \approx 1.9$) reported by the ladder D$\Gamma$A approach in the attractive Hubbard model~\cite{DGAattractive}. Instead, our result is much closer to the theoretical prediction for the Heisenberg universality class ($\gamma = 1.4$), consistent with other diagrammatic extensions of DMFT~\cite{DGA2,DFandDMFTTN}.
In Fig.~\ref{Fig3}, we also directly compare the data from Ref.~\cite{DFandDMFTTN}, demonstrating good agreement with the dual fermion method. Moreover, our data extend deeper into the critical regime, capturing a clean power-law behavior with a single exponent and without requiring the subleading corrections argued to be necessary in Ref.~\cite{DGAattractive}.
Several additional observables including the specific heat and spectral function have also been computed within our variational‐perturbation framework. The corresponding results, presented in the Supplementary Information~\cite{Supplementaryinfo}, further confirm the robust corrections to DMFT afforded by our approach.

\noindent \textbf{Conclusions.} 
We have developed a systematic variational perturbative expansion around the DMFT solution to study the Néel transition in the half-filled 3D cubic Hubbard model for
$U<15$. Through the convergence behavior of diagMC, we accurately estimated the critical temperature, which recovers DMFT in the weak-coupling limit and revealing significant corrections at intermediate and strong coupling. Analysis of critical exponents shows that the mean-field behavior of DMFT crosses over to the Heisenberg universality class. This approach is far less computationally demanding than other DMFT extensions and is readily applicable to second-order phase transitions in correlated systems, such as unconventional superconductivity in cluster-DMFT, and holds promise for realistic materials studies.

\smallskip

\noindent \textbf{Acknowledgements.} We acknowledge the support of NSF DMR-2233892, NSF OAC-2311557, and grant from the Simons Foundation (SFI-MPS-NFS-00006741-06).

The data that support the findings of this article are openly available~\cite{Repo2}.
\bibliography{main}

\newpage
\appendix
\onecolumngrid
\section{END MATTER}
\twocolumngrid
\noindent \textbf{Details of Calculation.}
In the DMFT calculation, we use hybridization expansion continuous-time quantum Monte Carlo(CT-QMC)~\cite{ctqmc} for $U<8$ and a more efficient bold-equivalent impurity solver~\cite{boldsolver,boldsolver2} for $U \geq 8$. 
The maximum entropy method (MEM) and Padé approximation are used to obtain the real-axis spectral function $\mathcal{A}_{\textbf{k}}(\omega)=-\frac{1}{\pi} \mathrm{Im}(G_{\textbf{k}}(\omega))$.
The validity of DMFT spectra from analytical continuation is verified through the real-axis version of bold impurity solver, which is benchmarked using numerical normalization group at low temperature.

The self-energies in the expansion are evaluated using numerical convolution for low-orders ($n<4$), to ensure high precision, and Markov chain Monte-Carlo (MCMC) for higher orders.
The MC sampling is performed in momentum and imaginary time basis, and the self-energy is projected onto the following space-time basis
$\Sigma_\vk(\tau) = \sum_{l,\vec{n}}u_l(\tau)v_{\vec{n}}(\textbf{k}) a_{l,\vec{n}}$,
defined on a k-grid with various density up to $20\times20\times20$, where $u_l(\tau)$ are the functions constructed via singular value decomposition of the Fermionic kernel~\cite{SVDBasis}, 
and the momentum basis $v_{\textbf{n}}(\textbf{k})$ is similar to Fourier basis with a long-wavelength cutoff. 
The coefficients $a_{l,\vec{n}}$ are determined through the MC sampling process.

To calculate observables, we sample Feynman diagrams for the self-energy $\Sigma_\vk(i\omega)$, which determines the Green's function $G_\vk(i\omega)$ through the Dyson equation. The staggered local magnetization is then computed by
$
  m=\langle \hat{n}_{i \uparrow}-\hat{n}_{i \downarrow} \rangle=\frac{1}{N_k \beta}\sum_{i\omega_n,\textbf{k}} [G_{\textbf{k} \uparrow}(i\omega_n)-G_{\textbf{k} \downarrow}(i\omega_n)],
$
and the antiferromagnetic susceptibility is evaluated by introducing small staggered magnetic field $H_{(\pi,\pi,\pi)}$, and calculating the response $\chi_{(\pi,\pi,\pi)}=\Delta m_{(\pi,\pi,\pi)}/\Delta H_{(\pi,\pi,\pi)}$.
The total energy is calculated using the Migdal-Galitskii formula 
$E=\mathrm{Tr}(H_0G)+\frac{1}{2}\mathrm{Tr}(\Sigma G)$,
and the specific heat is obtained by taking the numeric derivative $C_v= dE/dT$. 

\noindent \textbf{Details of Variational Perturbation.}
Since our action depends on variational parameter $\alpha$, variational-independent observables are determined at $\alpha_{\mathrm{opt}}$, chosen to ensure the best possible convergence of the expansion. Further details regarding the calculation of the self-energy and the principle of optimal convergence are also provided in the Supplementary Information~\cite{Supplementaryinfo}. The results for the half-filled system are presented in the main text; those for the doped Hubbard model appear in the Supplementary Information~\cite{Supplementaryinfo}.

To obtain $m(T)$ we first need to search for optimal variational parameter $\alpha_{\mathrm{opt}}\in[0,1]$, which is obtained by the principal of minimal sensitivity. 
When exchange splitting $\alpha$ is too small (large), the magnetization is increasing (decreasing) with perturbative order. 
However, at the optimal exchange splitting $\alpha_{\mathrm{opt}}$ the magnetization is essentially independent of perturbative order.

\begin{figure}[H]
  \subfigure{\includegraphics[height=0.75\textheight, trim=40mm 0mm 30mm 0mm, clip]{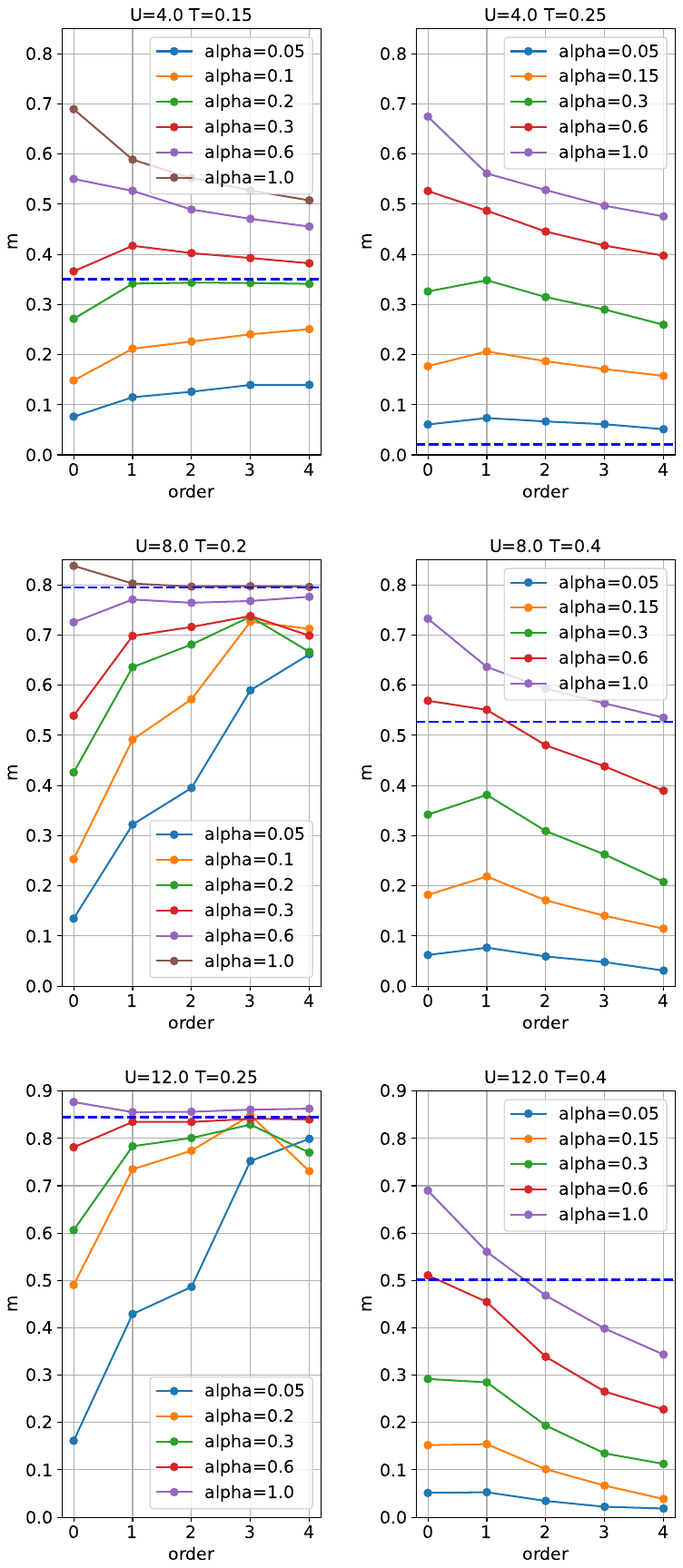}}
  \caption{Magnetization of 3D cubic Hubbard model at half-filling under different values of interactions, temperatures, orders, and variational parameters. 
  The blue dashed horizontal lines indicate the magnetization calculated from DMFT. The result of perturbation strongly relies on the starting point of perturbation.
  The points in the first column are all antiferromagnetic, while the points in the second column are all paramagnetic.  }
\label{mag_alpha_n}
\end{figure}

The magnetization after variational perturbation at different values of interaction and temperature is displayed in Fig.~\ref{mag_alpha_n}.
At $U=4$ and $T=0.15$, the optimal convergence is reached at around $\alpha=0.2$, since the magnetization changes only insignificantly with the perturbation order, after the first order. The finite magnetization is stable, and  is quite close to DMFT result, as expected for weak interaction.
In the paramagnetic phase of the same interaction strength (example $T=0.25$ in Fig.~\ref{mag_alpha_n}), the magnetization decreases with order for any positive $\alpha$, thus the magnetization vanishes.
In this weakly correlated regime, perturbing with a non-optimal $\alpha$ requires more orders, but the magnetization converges monotonically (except for the first order) to the same value.

At the point of highest $T_c$, and moderate correlation strength $U=8$, the convergence in the antiferromagnetic phase becomes substantially more challenging.
In the ordered phase, magnetization does not converge for all given $\alpha$'s (it diverges for $\alpha$'s far from optimum), however, we can still find a value of $\alpha_{opt}$, such that the convergence is rapid ($\alpha_{opt}\approx 1$ at $T=0.2$). Such large values of exchange splitting are of course expected in the antiferromagnetic states of correlated regime.
Interestingly, the convergence properties on the paramagnetic side are still as simple as in the weakly correlated regime, even at temperatures at which the DMFT predicts the long ranged ordered state. 
Even at $U=12$, the convergence in the ordered phase can still be achieved with large $\alpha\approx 1$. Most importantly, on the paramagnetic side the perturbation remains monotonic and straightforward to interpret, very similar to $U=4$.

Despite the difficulty in precisely determining the magnetization in the ordered state at $U=12$ (i.e., in the strongly correlated regime), it remains straightforward to assess whether $SU(2)$ symmetry is broken or restored with increasing perturbation order. 
By choosing a very small $\alpha$, the Néel temperature can be accurately determined. Using this approach, we obtained $T_c$ across the full correlation range, from the Slater to the Mott-Heisenberg limit.

\begin{figure}[htbp]
\centering
\hspace*{-0.8cm}
  \includegraphics[height=0.23\textheight, trim=0mm 0mm 0mm 0mm, clip]{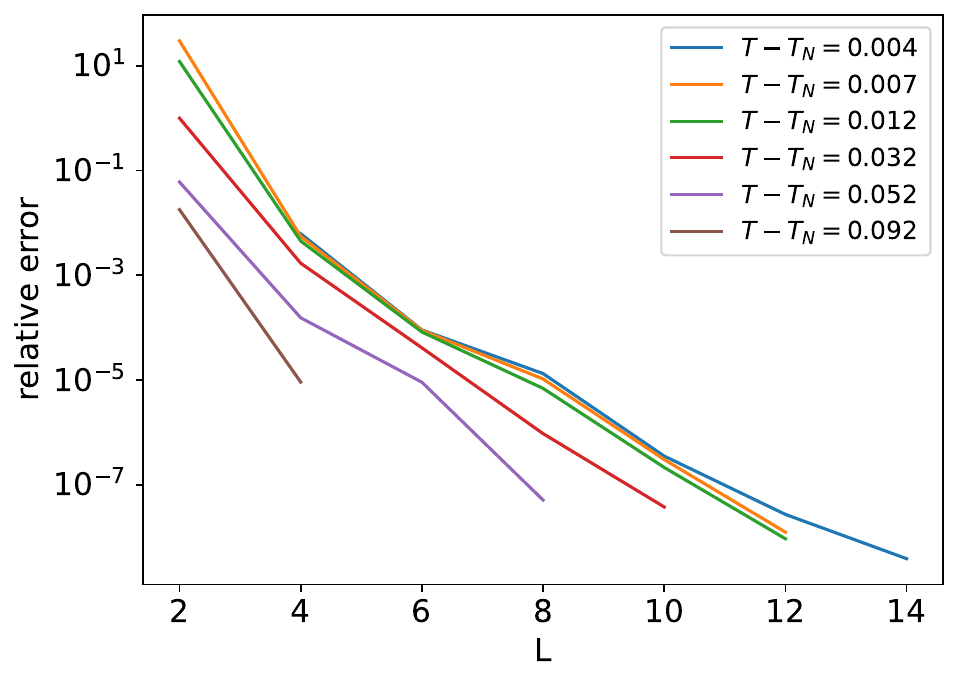}
  \caption{Relative error of $\chi_{AF}(L)$ at $U=11$ and various temperatures. As the temperature decreases, larger system sizes are required to reach equally high precision.}
\label{finite-size analysis}
\end{figure}

\noindent \textbf{Finite-size Analysis.}
To confirm the critical behavior, it is essential to cleanly separate finite-size effects from the genuine critical scaling. To this end, we increased the momentum lattice size up to $20\times 20\times 20$, and examined the relative error of $\chi$ as a function of lattice size $L\le 20$, defined as
$$|\chi_{AF}(L)-\chi_{AF}(L_{max}=20)|/\chi_{AF}(L_{max}=20).$$
While our method is formulated directly in the thermodynamic limit, the discretization of the momentum basis introduces a numerical approximation that requires careful verification.
In Fig.~\ref{finite-size analysis}, we present this relative error across different temperatures. As expected, the error increases as the temperature approaches the Néel temperature $T_N$, reflecting the growth of the correlation length and the corresponding need for larger lattices to maintain the same level of precision. For reference, Ref.~\cite{DFandDMFTTN} reports a correlation length of approximately $\xi \approx 3$ at $U = 11$ and $T - T_N = 0.01$. Under the same conditions, our relative error is approximately $10^{-4}$ for a linear lattice size of $L=6 \approx 2\xi$, consistent with their estimation. For lattices with $L > 14$, the relative precision of our data improves to better than $10^{-7}$.

\clearpage
\onecolumngrid

\section{Supplementary Information}

\twocolumngrid
\renewcommand{\thefigure}{S\arabic{figure}}
\setcounter{figure}{0}
\renewcommand{\theequation}{S\arabic{equation}}
\setcounter{equation}{0}


\subsection{Diagrammatics of Perturbative Expansion}

This section is dedicated to illustrating how perturbation theory is formulated diagrammatically.
As addressed before, in action Eq.2 (of the main text), we name the second term as `Hubbard U term' and the last term as `counter term'.
In the reorganized action Eq.2 the `unperturbed propagator' appears, which is defined as:
\begin{align}
  &G^0_{\textbf{k} \sigma}(i \omega_n) =(i\omega_n +\mu -\epsilon_{\textbf{k}} -\Sigma^{\mathrm{imp}}-\frac{\alpha}{2} U p_i \sigma_z)^{-1}\nonumber\\
  &=\left[
  \begin{array}{cc}
  i\omega_n +\mu  -\Sigma^{\mathrm{imp}}  -\frac{\alpha}{2}  U \sigma_z&  -\epsilon_{\textbf{k}} \\
  -\epsilon_{\textbf{k}}&  i\omega_n +\mu  -\Sigma^{\mathrm{imp}}+\frac{\alpha}{2}  U \sigma_z
  \end{array}
  \right]^{-1}
  \label{G0}
\end{align}
In addition, we also define the DMFT propagator $G^{\mathrm{DMFT}}$ by:
\begin{align}
  G^{\mathrm{DMFT}}_{\textbf{k}\sigma}(i\omega_n) &=(i\omega_n +\mu -\epsilon_{\textbf{k}} -\Sigma^{\mathrm{imp}}_{\sigma} (i\omega_n))^{-1}\nonumber\\
  &=\left[
  \begin{array}{cc}
  i\omega_n +\mu  -\Sigma^{\mathrm{imp}A}_{ \sigma}  &  -\epsilon_{\textbf{k}} \\
  -\epsilon_{\textbf{k}}&  i\omega_n +\mu  -\Sigma^{\mathrm{imp}B}_{\sigma}
  \end{array}
  \right]^{-1}
\end{align}

To investigate the antiferromagnetic phase of the system, Green's functions are defined on a doubled unit cell with two sites $A$ and $B$, while the $\textbf{k}$ is defined in the corresponding reduced Brillouin zone.
The four elements in the matrix represent $AA$, $AB$, $BA$, and $BB$ components of the Green's function. 
According to the DMFT approximation, the impurity Green's function is the local part of DMFT Green's function:
\begin{align}
  G^{\mathrm{imp}AA/BB}_{\sigma}=\sum_{\textbf{k}}G^{\mathrm{DMFT}AA/BB}_{\textbf{k}\sigma}  
\end{align}

\begin{figure}[htbp]
  \subfigure{\includegraphics[height=0.35\textheight, trim=0mm 0mm 0mm 0mm, clip]{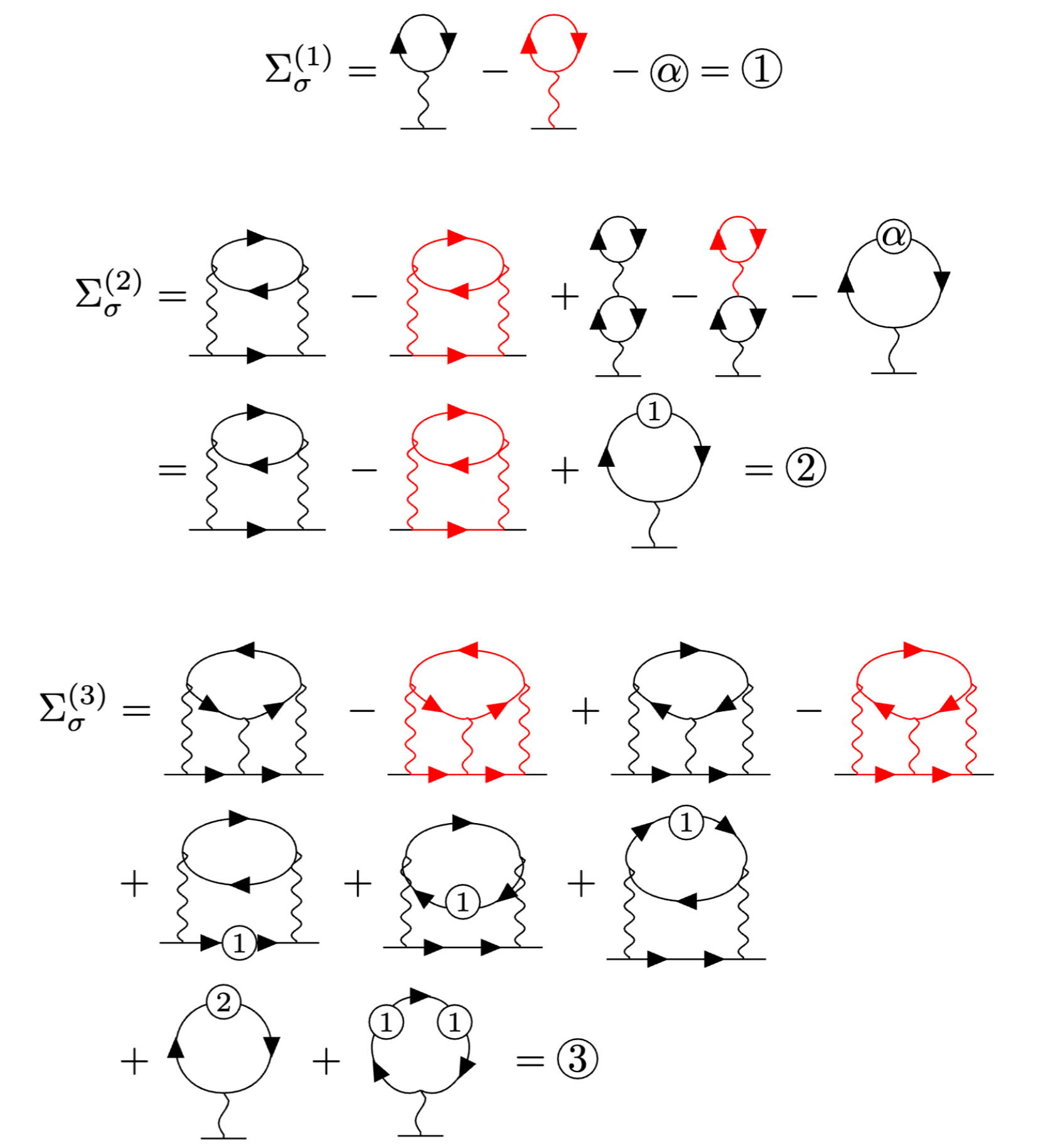}}
  \caption{Diagrammatic expansion of the first three orders.
  The circled $\alpha$ represents the variational splitting term $\alpha U p_i \sigma_z/2$.
  Black diagrams are constructed using the propagator $G^0$ with Hubbard $U$, while red diagrams are spin-averaged diagrams constructed from $G^{\mathrm{imp}}_{\sigma}$.
  }
\label{Diagrammatics}
\end{figure}

The diagrammatic expansion of the first three orders is shown in Fig.~\ref{Diagrammatics}.
In our notation, black diagrams are constructed using the propagator $G^0$ in Eq.\ref{G0} with Hubbard $U$, and red diagrams are part of $\Sigma^{\mathrm{imp}}$, which are spin-averaged diagrams constructed from $G^{\mathrm{imp}}_{\sigma}$.
In the first order, the self-energy simply contains the conventional `tadpole' diagram from the Hubbard $U$ term and another two terms from the counter term.
For simplicity, we refer the splitting term $\xi \alpha U p_i \sigma^z$ as circled $\alpha$, and the entire $n$th order self-energy as a circled $n$, as shown in Fig.~\ref{Diagrammatics}.

Similar to the first order, the second-order self-energy includes a skeleton diagram arising from the Hubbard $U$ term and a corresponding second-order diagram from the counter term $\Sigma^{\mathrm{imp}}$.
Additionally, a non-skeleton diagram, the `tower', should also be included in the self-energy.
Moreover, second-order diagrams can be generated by inserting a first-order counter term, which is part of $\Sigma^{\mathrm{imp}}$, into another first-order diagram, formed by $G^0$ and the Hubbard $U$ term.
Similarly, the constant splitting term inserted in a `tadpole' diagram should also be a part of second-order self-energy, represented by the last term.
Such insertions are the main difference between our perturbation expansion and the typical perturbation theory with only one interaction term.

Once we calculate the first-order self-energy, all non-skeleton diagrams of the second order can be combined into a single diagram, which is a `tadpole' with the entire first-order self-energy inserted, as shown in the third line of Fig.~\ref{Diagrammatics}.
Higher-order diagrammatic expansion can be significantly simplified using this trick. Instead of listing all diagrams explicitly, the $n$th order diagrams can be simply expressed recursively in two parts: 
(a) skeleton diagrams constructed using $G^0$ and Hubbard $U$, along with corresponding counter diagrams from $\Sigma^{\mathrm{imp}}_{PM}$, constructed from $G^{\mathrm{imp}}_{\sigma}$;
(b) All possible low-order self-energy insertions into lower-order skeleton diagrams constructed using $G^0$ and Hubbard $U$, which gives the same total order $n$.
By recursively using low-order self-energies, the calculation of high-order self-energy is just an evaluation of skeleton diagrams, as the third-order diagrams shown in Fig.~\ref{Diagrammatics}.

\subsection{Detailed Procedure of VDMC}
Before proceeding, we summarize the detailed procedure of our algorithm:

\begin{enumerate}
    \item Perform a standard DMFT calculation in the magnetically ordered state to obtain the impurity Green’s function $G^{\mathrm{imp}}_{\sigma}(i\omega_n)$, the  spin-resolved impurity self-energy $\Sigma^{\mathrm{imp}}_{\sigma}(i\omega_n)$ and the spin-averaged impurity self-energy $\Sigma^{\mathrm{imp}}(i\omega_n)$.
    \item Using $G^{\mathrm{imp}}_{\sigma}(i\omega_n)$, evaluate all local skeleton diagrams of order $n$; denote the result by $\Sigma^{\mathrm{imp}(n)}_{\sigma}$. The spin-averaged counterpart is $\Sigma^{\mathrm{imp}(n)}=(\Sigma^{\mathrm{imp}(n)}_{\uparrow}+\Sigma^{\mathrm{imp}(n)}_{\downarrow})/2$. These quantities serve as the counterterms in the perturbation series.
    \item Search for the optimal exchange-splitting parameter $\alpha_{\mathrm{opt}}$. For each trial $\alpha$ perform the following:
    \begin{enumerate}
        \item Build the unperturbed Green’s function: $$G^0_{\sigma}(\alpha,i\omega_n,\textbf{k})=[i\omega+\mu-\epsilon_{\textbf{k}}-\Sigma^{\mathrm{imp}}(i\omega_n)-\sigma^{z}_{\sigma}p_i/2]^{-1},$$ where $\sigma^z$ is the Pauli matrix, $\Sigma^{\mathrm{imp}}(i\omega_n)$ is the spin-averaged impurity self-energy from step 1, $p_i=(-1)^{i}\alpha U$
        alternates on the two sublattices denoted by $i=0,1$, $U$ is the Hubbard interaction. 
        \item Generate and evaluate all diagrams of the order \(\xi^{n}\) from the action (Eq. 2, main text) using the building blocks
        \(\{G^{0}_{\sigma}, U, \alpha\}\) and the counterterms from step 2. The diagrams are obtained by following the standard diagrammatics rules, displayed in Fig.~\ref{Diagrammatics}. 
        Sum them to obtain \(\Sigma(\alpha,n)\) for all orders \(n \le N\).
        \item Calculate the best estimate for the physical Green's functions $G(\alpha,N)$ using the Dyson's equation: $$[G(\alpha,N)]^{-1}=[G^0(\alpha)]^{-1}-\sum^{N}_{n=1} \Sigma(\alpha,n).$$ Here we droped the momentum $\vk$, frequency $i\omega_n$ and spin $\sigma$ for brevity.
        \item Compute the desired quantity 
        \(\mathcal{O}(\alpha,N)\) from 
        \(G(\alpha,N)\); for instance,
        the magnetization is $m=T\sum_{i\omega_n,\textbf{k}}[G_{\uparrow}(\alpha,N,i\omega_n,\textbf{k})-G_{\downarrow}(\alpha,N,i\omega_n,\textbf{k})]/2$).
        \item Evaluate the convergence of 
        \(\mathcal{O}(\alpha,N)\) with increasing \(N\).  
        Define \(\alpha_{\text{opt}}\) as the value that minimizes the standard deviation
        \(\sigma_{N}\!\bigl[\mathcal{O}(\alpha,N)\bigr]\) over the last few perturbative orders.
    \end{enumerate}
    \item The best estimate of the observable $\mathcal{O}$ is obtained from results of the two highest orders at $\alpha_{\mathrm{opt}}$. 
\end{enumerate}

\subsection{Estimation of Magnetization}
As explained in the main text, the Néel temperature can also be calculated with higher precision through extrapolation of the critical behavior of magnetization.
In our approach, we calculate the standard deviation and find the $\alpha_{opt}$ which minimizes it.
\begin{figure}[t]
  \subfigure{\includegraphics[height=0.65\textheight, trim=0mm 0mm 0mm 0mm, clip]{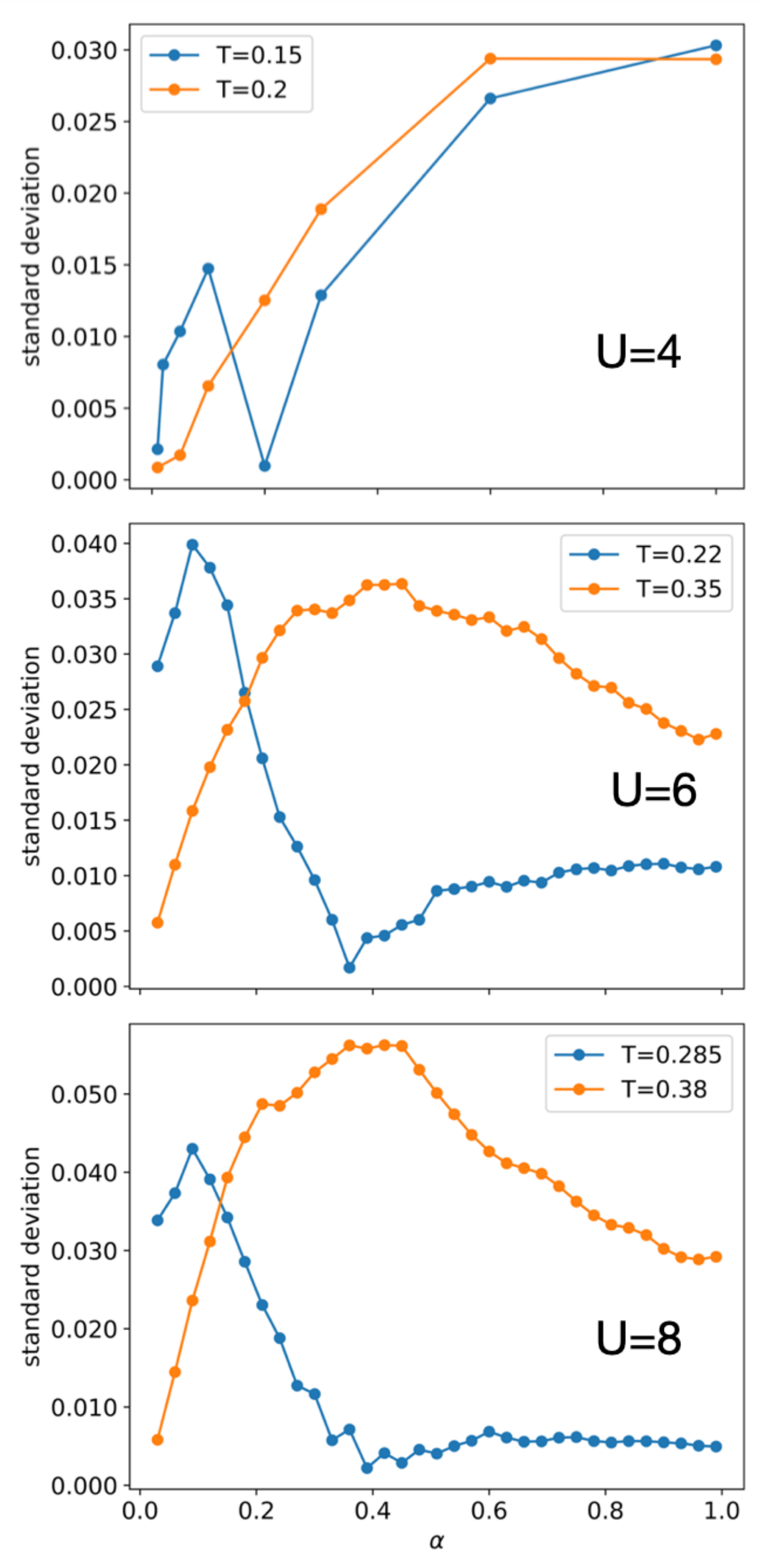}}
  \caption{Standard deviation as a function of $\alpha$. 
  In antiferromagnetic states (blue curves), there is a minimum other than 0. However, such points do not exist in paramagnetic states (orange curves).}
\label{stddev}
\end{figure}
Fig.~\ref{stddev} shows the $\alpha$-dependence of the standard deviation defined above. 
Generally speaking, $\alpha=0$ must have zero standard deviation since we are stuck in paramagnetic state.
However, this trivial point is not always the desired physical minimum point. As shown in Fig.~\ref{stddev}, there exists another physical local minimum at finite splittings for ordered states. 
For example, at $U=4$ and $T=0.15$, the physical minimal standard deviation is achieved at $\alpha_{opt}=0.2$ instead of $\alpha=0$, which corresponds to the optimal convergence displayed in Fig.~\ref{stddev}.
As the temperature approaches the Néel temperature, the optimal alpha will decrease and finally be zero at the phase transition point.
This explains why minimums at non-zero $\alpha$s do not appear for $T>T_c$, and $\alpha=0$ is optimal for paramagnetic states.

Unfortunately, our criterion for optimal alpha fails once the system enters the Mott insulating state of DMFT. 
As shown in Fig.~\ref{stddev}, the standard deviation at large $\alpha$ values keeps decreasing as the interaction gets stronger.
At $U=8$ and $T=0.285$, the standard deviation at $\alpha>0.6$ is just slightly higher than the standard deviation at $\alpha_{opt}$.
In Mott insulating state, minima of standard deviation are always reached at very large $\alpha$s in the antiferromagnetic phase, even near the Néel temperature.
Since the resulting magnetization is not so straightforward to determine, we prefer to avoid this regime in this work. It will require higher order in perturbation
theory before convergence can be achieved.

\subsection{Néel Temperature in the Doped Regime}
\begin{figure}[htbp]
  \centering
  \hspace*{-0.7cm}
  \includegraphics[height=0.3\textheight, trim=0mm 0mm 0mm 0mm, clip]{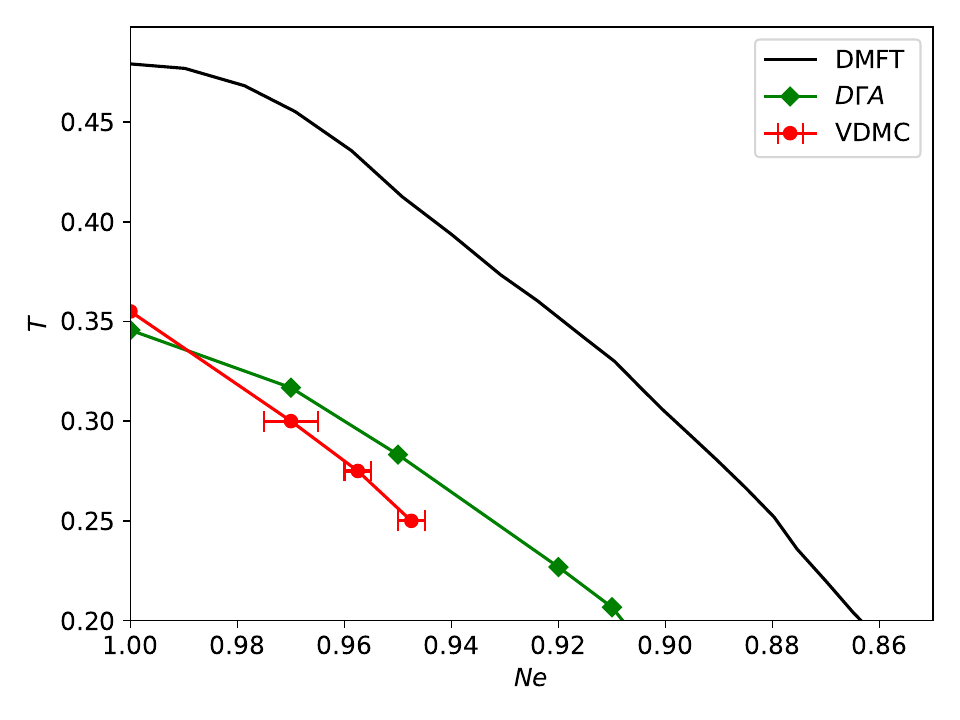}
\caption{Phase diagram in the doped regime at $U=10$, compared with DMFT\cite{DopedDMFT} and D$\Gamma$A \cite{dopedDGA} (D$\Gamma$A value is at $U=4 \sqrt{6}\approx 9.8$)}
  \label{dopedTN}
\end{figure}

To show that our method also works away from half-filling, we examined the more challenging case of finite doping. The calculation reproduces the rapid suppression of the Néel temperature with doping and agrees quantitatively with the dynamical vertex approximation (D$\Gamma$A) \cite{dopedDGA}.

For the doped regime we perform the expansion at fixed electron density $N_e$: we first evaluate the DMFT self-energy at that density, then adjust the chemical potential $\mu$ in the perturbed Green’s function so that every perturbative order also yields density $N_e$. The resulting shift of the chemical potential $\mu$ is included in the counterterm.

Figure~\ref{dopedTN} presents the Néel temperature at 
$U=10$ for several dopings, together with DMFT \cite{DopedDMFT} and D$\Gamma$A \cite{dopedDGA} results. The VDMC points follow the D$\Gamma$A curve closely and lie well below the DMFT values.

Although expanding at fixed density is the most straightforward option, it is not unique; convergence could be accelerated by introducing additional variational freedom. In particular, even a constant shift of the action can improve convergence in the paramagnetic phase away from half-filling \cite{alpha_shift}.

\begin{figure}[htbp]
\centering
  \includegraphics[height=0.3\textheight, trim=0mm 0mm 0mm 0mm, clip]{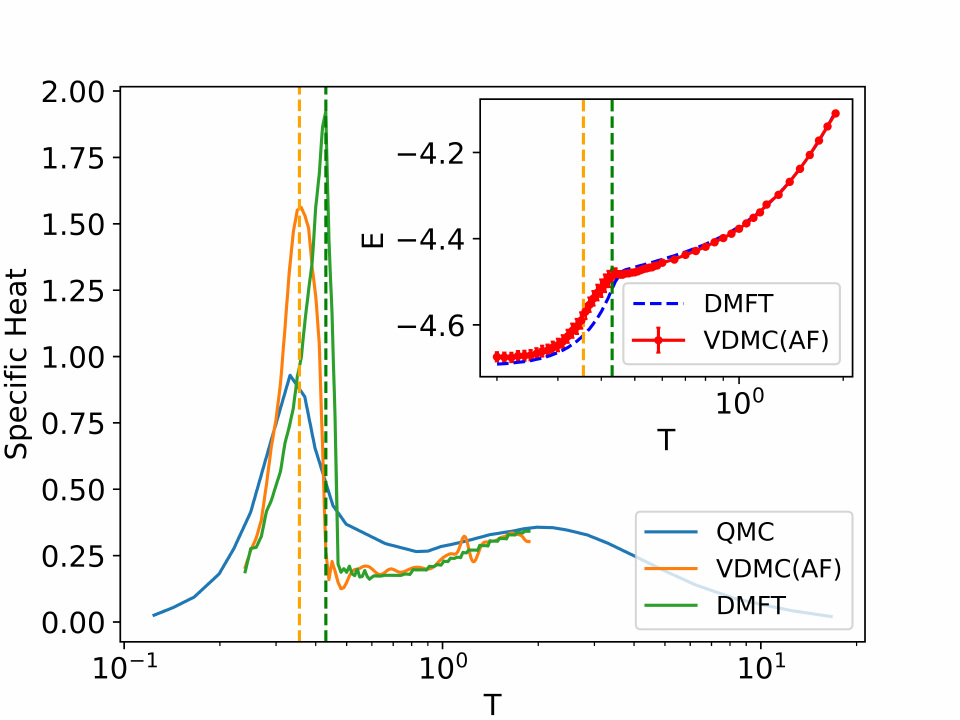}
  \caption{Specific heat from VDMC(AF), DMFT and QMC\cite{ThermoQMC} as a function of temperature at $U=8$. 
  The dashed vertical lines indicate the positions of peaks of the specific heat curves: $T=0.355$ for VDMC(AF) and $T=0.43$ for DMFT.
  Inset: Total energy density from DMFT and VDMC(AF). The dashed lines have the same position as the main figure.}
\label{Cv}
\end{figure}

\subsection{Thermodynamics from VDMC}
Across the Néel transition, the thermodynamic properties also serve as an indicator of the universality class.
Specifically, universality classes with critical exponent $\alpha>0$ exhibit a diverging peak in specific heat ($c_V$), 
while universality classes with $\alpha<0$ show a finite peak or just a kink across the second-order phase transition.
Although several numerical methods show a finite peak\cite{AFQMCTN,ThermoQMC} or just a kink\cite{Cvkink} for $c_V$ in the 3D Hubbard model,
DMFT approximation exhibits an abrupt jump when approaching the Néel transition from the paramagnetic side and a sharp peak across the phase transition, as shown
in Fig.~\ref{Cv} for $U=8$.

\begin{figure}[htbp]
\centering
\hspace*{-0.5cm}
 \includegraphics[scale=0.43, trim=0mm 0mm 0mm 0mm, clip]{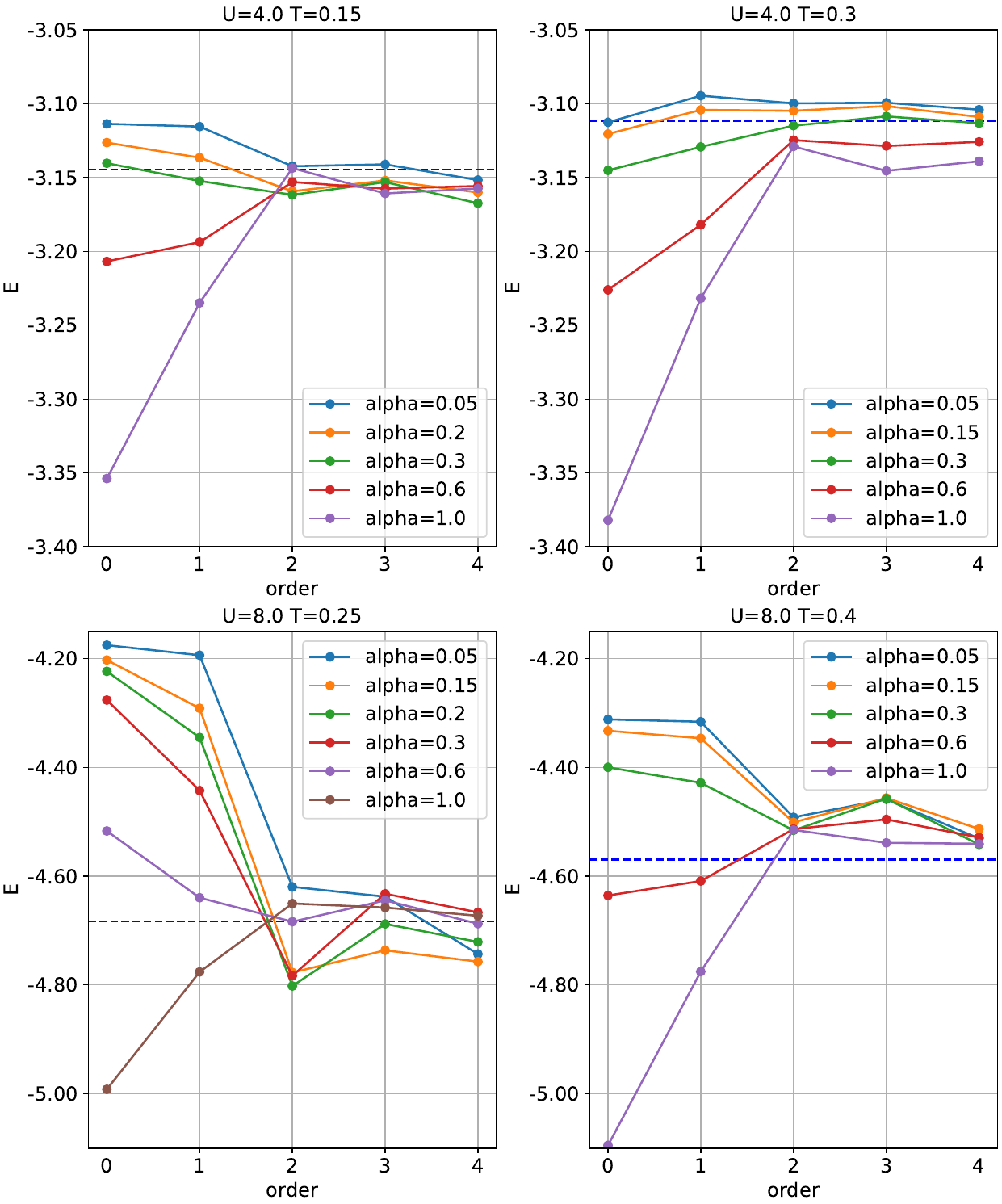}

    \caption{Total energy of 3D cubic Hubbard model at half-
  filling under different values of interactions, temperatures, orders and variational parameters. The blue dashed horizontal
  lines indicate the total energy calculated from DMFT.}
  \label{E_alpha_n}

\end{figure}

Furthermore, the total energy obtained from DMFT shows a kink across the transition, indicating $dE/dT$ may not be continuous.
We attribute these effects to the mean-field nature of DMFT, which yields a critical exponent for the specific heat of $\alpha=0$, 
which can still lead to logarithmic divergence in the specific heat and a discontinuous of $dE/dT$.

For temperatures above the DMFT Néel temperature, our perturbative correction reduces the total energy slightly as compared to DMFT, but this has a very limited impact on the specific heat. Strong corrections of $c_V$ occur in the range $T_N<T<T^{\mathrm{DMFT}}_N$, where the perturbatively  corrected total energy exceeds that obtained from DMFT (see inset of Fig.~\ref{Cv}). This correction smoothens the steep jump in DMFT specific heat, which is now on the paramagnetic side of the true Néel transition, and the rise of $c_V$ is interpreted to be due to critical fluctuations near the 3D phase transition, absent in infinite dimensions. The diverging peak of DMFT is thus replaced by a broader peak, 
which is consistent with the properties of the Heisenberg universality class with negative critical exponent $\alpha=-0.12$.
Moreover, the peak shifts from the DMFT Néel temperature $T^{\mathrm{DMFT}}_N=0.43$ to the true Néel temperature $T=0.355$, which is very close to the position of the peak found by QMC\cite{ThermoQMC} method, which we also reproduce in  Fig.~\ref{Cv}). Note that $T_N$ estimated from $c_V$ and from magnetization match well, proving internal consistency of the method.
By employing a low-cost perturbative expansion with only a few orders, we capture several correct features of the specific heat curve, 
including the peak position and a qualitatively correct critical behavior,  which demonstrate the potential power of our method.
Further details of variational perturbation of total energy are discussed in supplementary information.
The convergence behavior of total energy is displayed in Fig\ref{E_alpha_n}. 
Similar to magnetization, the zeroth-order total energy strongly depends on the choice of $\alpha$. 
However, if temperature is outside the window $T_N<T<T^{\mathrm{DMFT}}_N$, the total energy for any $\alpha$ values converges to the value very close to the DMFT solution just after the second-order perturbation. After the second order, total energy starts to fluctuate, and the fourth order does not reach full convergence.
At $U=4$, where $\alpha_{opt}\sim 0.2$ for $T=0.15$ and $\alpha_{opt} \sim 0$ for $T=0.3$, the size of perturbation correction is about $0.01t$, 
which is a small error bar, since the total energy changes from $-4.7t$ to $-4.4t$ as the temperature increases from $0.1t$ to $t$.
At $U=8$ and $T=0.25$, the total energy from optimal $\alpha_{opt} \sim 0.6$ still matches the DMFT result at the fourth order. 
However, the total energies show $\sim0.05t$ positive corrections to the DMFT result at $T=0.4$, which is between $T_N$ and $T^{\mathrm{DMFT}}_N$.
Since this correction can be seen from all $\alpha$s, it should be considered a solid correction independent of the variational setting of the perturbation.
As shown in Fig.~\ref{Cv} in the main text, most significant corrections to the total energy occur below $T^{\mathrm{DMFT}}_N$ but above $T_N$.

\begin{figure*}[htbp]
  \centering
  \includegraphics[scale=0.9, trim=0mm 95mm 0mm 100mm, clip]{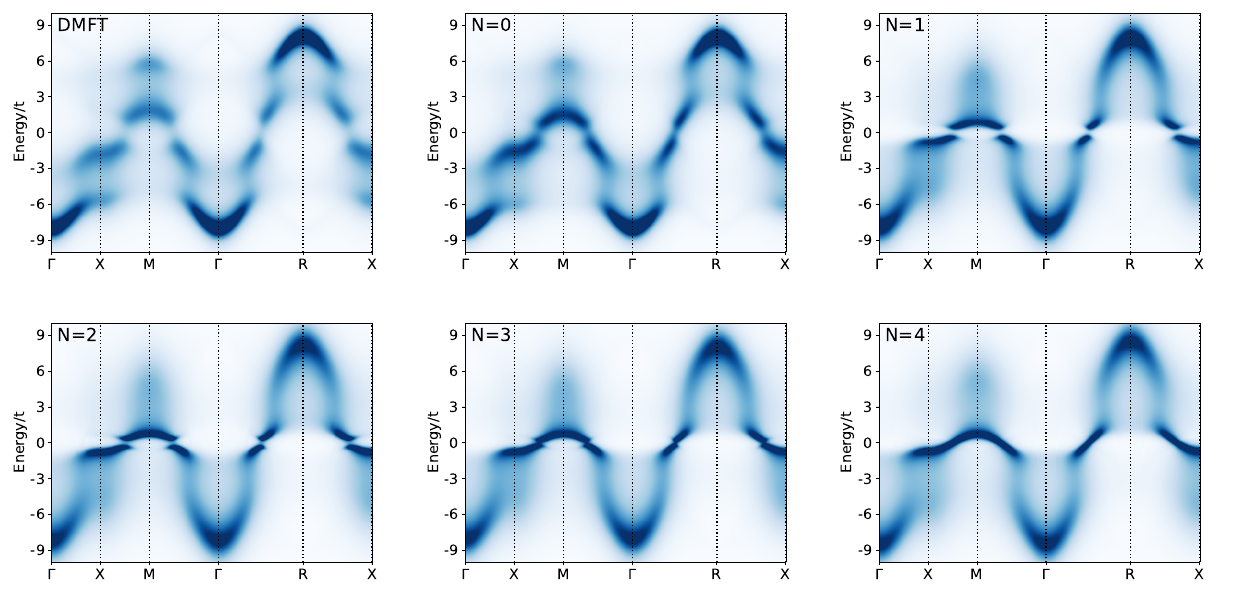}
  \caption{Spectral functions at $U=8$, $T=0.41$ from DMFT and perturbation up to order four. $\alpha=0.3$ is used in perturbation results. The k-path is built on a standard cubic Brillouin zone with high-symmetry points: $\Gamma=[000], X=[010], M=[110], R=[111]$.}
  \label{Fig5}
\end{figure*}
\subsection{Correction of Spectra}
The spectral functions provide the most direct evidence of how an antiferromagnetic state in DMFT is transformed into a paramagnetic state in our scheme.
Since the Néel temperature has been significantly reduced in the intermediate coupling, we expect the DMFT gap size to be reduced below $T_N$, and even eliminated between $T_N$ and $T^{\mathrm{DMFT}}_N$.
Fig.~\ref{Fig5} shows how spectral functions are modified by perturbation order by order at $U=8$, $T=0.41$, with $\alpha=0.3$.
At this intermediate temperature, most nonlocal extensions of DMFT give a paramagnetic metal rather than the antiferromagnetic insulator seen in the DMFT solution. 
In DMFT, the quasi-particle band close to the Fermi level is split by a small Slater gap, and with Hubbard bands located at much higher energy.
The 0th-order starting point at $\alpha=0.3$ remains a gapped state, retaining most features of the DMFT spectrum with a reduced gap size around the Fermi level.
As the perturbation order increases, the main structure of the DMFT spectrum is still preserved.
However, rather than expanding the gap to recover the DMFT-like state, the gap size continues to shrink with each order of perturbation.
In the fourth order, the gap is eliminated, leaving a metallic quasi-particle band. 
Moreover, although $\alpha$ is a tunable parameter and $\alpha=0.3$ gives rapid convergence, starting with different $\alpha$'s does not affect the conclusion that perturbation will eventually close the gap. The only issue is that more orders may be required to reach this paramagnetic metallic state when using larger $\alpha$'s.
Another antiferromagnetic example is discussed in the Supplementary Information.

\begin{figure}[H]
\includegraphics[scale=0.55, trim=0mm 0mm 0mm 0mm, clip]{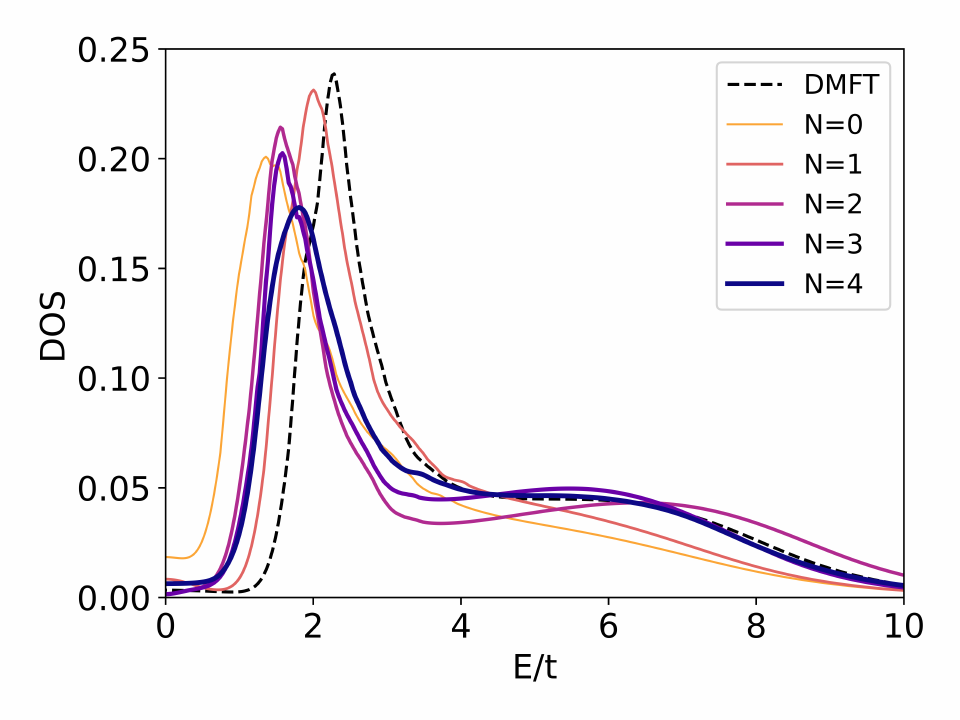}
  \caption{Density of states at $U=8$ and $T=0.25$ from DMFT and perturbation up to order four. $\alpha=0.3$ is used in perturbation results. 
  Results from higher orders are represented by darker and thicker solid lines. 
  The density of states above and below the Fermi surface is symmetric due to the particle-hole symmetry at half-filling.}
\label{DOS}
\end{figure}

Our perturbation scheme corrects the spectral function not only between $T_N$ and $T^{\mathrm{DMFT}}_N$ but also deep in the ordered state.
The local density of states at $U=8$, $T=0.25$ is displayed in Fig.~\ref{DOS}.
In this example, we start with $\alpha=0.3$, as it allows the spectra around the gap to converge rapidly.
As expected, the gap size and the density of states around the Fermi level are significantly modified, while the high-frequency spectrum is converged back to the DMFT solution.
In the DMFT solution, the gap size is around $3t$, but by tuning the splitting, our perturbation starts from a smaller gap $\sim t$.
After the first order, the gap expands to around $2t$ and shrinks to $\sim 1.8t$ at the second order. 
Despite some fluctuations outside the gap, the gap size stabilizes at around $1.8t$ at the higher orders of perturbation.

\end{document}